# Impact of Climate Change on Forests in India

N. H. Ravindranath[1], N.V. Joshi[1], R. Sukumar[1] and A. Saxena[2]

[1]Centre for Ecological Sciences, Indian Institute of Science, Bangalore – 560 012

[2]Forest Survey of India, Dehra Dun

ravi@ces.iisc.ernet.in

**Global assessments have shown that future climate change is likely to significantly impact forest ecosystems. The present study makes an assessment of the impact of projected climate change on forest ecosystems in India. This assessment is based on climate projections of Regional Climate Model of the Hadley Centre (HadRM3) using the A2 (740 ppm $CO_2$) and B2 (575 ppm $CO_2$) scenarios of Special Report on Emissions Scenarios and the BIOME4 vegetation response model. The main conclusion is that under the climate projection for the year 2085, 77% and 68% of the forested grids in India are likely to experience shift in forest types under A2 and B2 scenario, respectively. Indications are a shift towards wetter forest types in the northeastern region and drier forest types in the northwestern region in the absence of human influence. Increasing atmospheric $CO_2$ concentration and climate warming could also result in a doubling of net primary productivity under the A2 scenario and nearly 70% increase under the B2 scenario. The trends of impacts could be considered as robust but the magnitudes should be viewed with caution, due to the uncertainty in climate projections. Given the projected trends of likely impacts of climate change on forest ecosystems, it is important to incorporate climate change consideration in forest sector long-term planning process.**

**Introduction**

Climate is probably the most important determinant of vegetation patterns globally and has significant influence on the distribution, structure and ecology of forests[1]. Several climate-vegetation studies have shown that certain climatic regimes are associated with particular plant communities or functional types[2,3,4,5]. It is therefore logical to assume that changes in climate would alter the configuration of forest ecosystems[6,7].

The Third Assessment Report of IPCC[8] concluded that recent modelling studies indicate that forest ecosystems could be seriously impacted by future climate change. Even with global warming of 1-2°C, much less than the most recent projections of warming during this century[9], most ecosystems and landscapes will be impacted through changes in species composition, productivity and biodiversity[10]. These have implications for the livelihoods of people who depend on forest resources for their livelihoods[11].

India is a mega-biodiversity country where forests account for about 20% (64 million ha) of the geographical area[12]. With nearly 200,000 villages classified as forest villages, there is obviously large dependence of communities on forest resources[13]. Thus it is very important to assess the likely impacts of projected climate change on forests and develop and implement adaptation strategies for both biodiversity conservation and the livelihoods of forest-dependent people.

Preliminary qualitative assessments of potential climate change impacts on forests in India[14,15] were based on earlier GCM (General Circulation Model) outputs of climate change[16,17] that have undergone considerable refinement. Following this there were two regional studies, the first pertaining to potential climate change impacts on forests in the northern state of Himachal Pradesh[18], and second in the Western Ghats[15]. These studies indicated moderate to large-scale shifts in vegetation types, with implications for forest dieback and biodiversity. The studies conducted in India so far have had several limitations, e.g. coarse resolution of the input data and model outputs due to the use of GCM scale grids, the use of earlier versions of the BIOME model that had limited capability in categorizing plant functional types, and the absence of any national level model-based assessment of climate impacts. A recent study[19] using BIOME3 model and climate change scenarios of HadCM2 projected large-scale shifts in areas under different vegetation types and an increase in NPP. As part of our ongoing efforts in refining our predictive capabilities, the present study assesses the potential impacts of future climate change on forest ecosystems at the national level based on RCM (Regional Climate Model) projections and a more advanced version of the BIOME model.

**Methods**

To assess the impact of climate change on the forests in India, changes that are expected to take place in location as well as the extent of the different types of forests in India were estimated. This was based on (i) data on the spatial distribution of forests across India, (ii) data on current climate consisting of the spatial distribution of climatic variables (rainfall, temperature, cloud cover), (iii) similar data for future climate, as projected by relatively high-resolution global/regional climate models for two different climate change scenarios, both baseline as well as projections available as 30-year long time series, (iv) data on water-holding capacity and depth of topsoil and subsoil at various locations in India and, most important, (iv) the model BIOME4[20] , which uses these data as input and suggests the vegetation types most likely to occur at the corresponding locations.

**Vegetation impact assessment model and data sources**

The assignment of the vegetation types was carried out using the BIOME4 model developed by J.O. Kaplan and I.C. Prentice at the Max-Planck Institute. BIOME4[20] is in a way a successor of the BIOME3 model[21] and works using a "coupled carbon and water flux scheme, which determines the seasonal maximum leaf area index (LAI) that maximizes NPP for any given PFT, based on a daily time step simulation of soil water balance and monthly process-based calculations of canopy conductance, photosynthesis, respiration and phenological state. The model is sensitive to $CO_2$ concentration because of the responses of NPP and stomatal conductance to $CO_2$ and the differential effects of $CO_2$ on the NPP of C3 and C4 plants[20]". The driver routine was modified for reading data specific to the locations (latitudes and longitudes) of interest to the present study. The climate-related input variables required by this model (for each of the locations, i.e. for a latitude-longitude grid) are the monthly mean temperature (°C) and precipitation (mm) and monthly sunshine hours (% of maximum). In addition, the model also requires data on the water holding capacity (WHC in mm) of the top 30 cm of the soil, WHC of the rest of the soil (next 120 cm), and the conductivity indices of water through the soil columns for these two layers. The model evaluates the net primary productivity (NPP) for various values of the leaf area index for 13 plant functional types (PFTs), and thus determines the maximum NPP value for each of them. Based on these, BIOME4 assigns each grid to one of the 28 different vegetation types.

*Observed climate data*

The global climate data set for a 10 minute x 10 minute grid available from the Climate Research Unit (CRU) of the University of East Anglia[22] was used in the present study. This represents the present climate, based on the data centered on the period 1960-1990. Monthly values of rainfall, and mean monthly values of temperature and cloud cover were available as a time series. These were used to obtain the monthly averages needed by the BIOME4 model.

*Climate projections from regional climate model (HadRM3) of the Hadley Center*

Among the various GCMs currently available, the Hadley Centre model (Had) was used for climate projections because this model provides regional-scale projections. Projected values of climate variables for the two future scenarios for the period 2071-2100 were provided by the Indian Institute for Tropical Meteorology (IITM), which had been obtained from the runs of HadRM3 model adapted for the Indian region. These were available for a 0.44 degree x 0.44 degree grid. The IITM also provided data for a control run, corresponding to the notional period 1960-1990. While the HadRM3 outputs contained data on precipitation as well as on maximum and minimum temperature, it recorded the percentage of cloud cover instead of percentage of sunshine hours needed by the BIOME4 model. The conversion of cloud cover to sunshine hours was carried out based on the approach of Doorenbas and Pruitt[23] and using piecewise linear transformation as described in Hulme et al.[24]. The climate variables for future scenarios were obtained using the method of anomalies[20]. Briefly, this involved computing the difference between the projected values for a scenario and the control run of the model, and adding this difference to the value corresponding to the current climate as obtained from the CRU 10 minute climatology. The present investigations have been carried out for the year 2085 for the A2 scenario (approx 740 ppm $CO_2$) as well as the moderate B2 scenario (575 ppm $CO_2$), while a value of 320 ppm of $CO_2$ was used for the present climate.

*Soil parameters*

The values used for the soil parameters were based on the FAO classification, and were extracted from the data file for a 30 minute x 30 minute global grid enclosed with the BIOME4 code.

*Forest types and density*

The locations of the grid points corresponding to the Indian region were determined from the digital forest map prepared by the Forest Survey of India (FSI). This was based on high-resolution mapping (2.5' by 2.5'), amounting to over 165,000 grids, covering all the states and union territories of India. Data on the presence/absence of one or more of the 22 forest types as well as of scrub vegetation is mapped. The record for each grid consisted of its location and classification (forest, non-forest/scrub). If it was forested, details of the vegetation type in the grid and the crown density (10-40%, 40-70% and above 70%) were also recorded. In all, over 35,000 grids were recorded as forested.

*Selection of grids*

Based on the high-resolution map of the FSI, a total of 1696 grids of HadRM3 were seen to fall in the Indian region. Similarly, 10864 grids of the CRU could be assigned to the Indian region. A total of 4,469 of the CRU grids contained forest as recorded by the FSI and accounted for over 35,000 of the FSI grids. We have used the high-resolution FSI grids for all the analysis carried out in the present investigation.

*Selection of emission scenarios*

Emission scenarios selected for climate projection are A2 and B2 based on IPCC-Special Report on Emission Scenarios[25]. The projected climate patterns were obtained using the method of anomalies[20], i.e., by taking the difference between the baseline value of HadRM3 and corresponding value under A2 (or B2) scenario, and adding this difference to the corresponding value for the grid obtained from the CRU data. Since the projected data was for the period 2071-2100, mean values are taken to represent the year 2085. The A2 scenario represents (globally) a growing human population, and slower and inequitable economic development whereby atmospheric $CO_2$ concentration is projected to double by 2050 and is likely to reach 740 ppm by 2085. In the B2 scenario that represents moderate population growth, intermediate levels of economic development, adoption of environmentally sound technologies and greater social equity, $CO_2$ concentration is projected to reach 575 ppm by 2085.

**Results and Discussion**

**Current and future climate patterns**

The mean annual precipitation over India as computed from the CRU data was seen to be about 1094 mm and the mean annual temperature was about 22.7°C. The projected climate (average for 2071-2100) for the more moderate B2 scenario is both wetter (an average increase of about 220 mm) and warmer (an average increase of about 2.9°C) compared to the HadRM3 baseline. The corresponding values of increase for the more extreme A2 scenario are about 300 mm and 4.2°C respectively. The mean annual precipitation for the projected values for B2 scenario turns out to be 1314 mm and the projected mean temperature is about 25.6°C. There is considerable geographical variation in the magnitude of changes for both temperature (Figure 1a) as well as rainfall (Figure 1b). Northwestern India is likely to become drier, while northeastern India is likely to become much wetter. The temperature increase in northwestern India is also much more than that in the northeast. Southern and southeastern parts of India are likely to experience only a moderate increase in temperature.

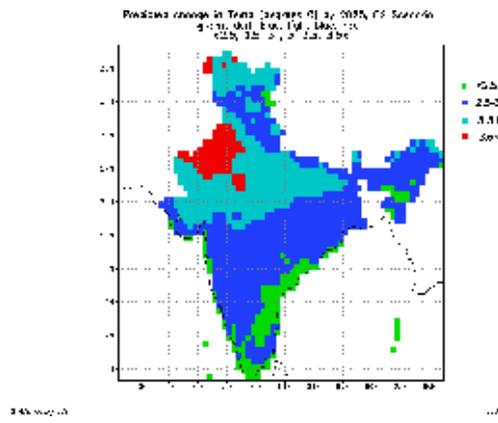 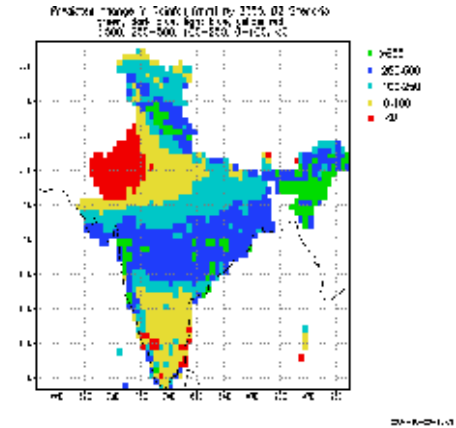

**Figure 1a.** Predicted change in temperature (°C) by 2085, B2 scenario

**Figure 1b.** Predicted change in rainfall (mm) by 2085, B2 scenario

*Changes in climate over forested areas*

As explained earlier, the high-resolution data from FSI can be used to map the location of various types of forests across India. The major forest types in India (those occupying 0.5% or more of the forested area) and their areas are given in Table 1. Forests in India are extremely diverse and very heterogeneous in nature, and it is difficult to classify them into a small number of categories. As a result, the pan-Indian 'Miscellaneous forest' category (with no dominant species) shows the highest (63%) proportion. The miscellaneous forest area occurs under all the forest types. The other two most dominant forest types are *Shorea robusta* or Sal (12%) in the eastern part of Central India and *Tectona grandis* or Teak (9.5%), spread across Central India as well as the Western Ghats in Southern India. The climate impact analysis is carried out for the FSI forest categories as well as the BIOME model vegetation types.

**Table 1.** Annual rainfall and temperature changes in the different forest types of India under B2 GHG scenario for the year 2085

| Forest type | Number of grids | % area | Mean annual rainfall (mm) | Change in rainfall (mm) | Mean temperature (°C) | Change in temperature (°C) |
|---|---|---|---|---|---|---|
| Fir | 290 | 0.82 | 730.1 | 221.6 | 9.5 | 3.0 |
| Blue-Pine (Kail) | 311 | 0.88 | 763.0 | 223.5 | 10.5 | 3.0 |
| Chir-pine | 791 | 2.25 | 1373.4 | 437.4 | 17.1 | 2.8 |
| Mixed conifer | 1071 | 3.04 | 930.1 | 375.9 | 9.3 | 3.0 |
| Hardwoods Conifers mix | 296 | 0.84 | 1560.7 | 585.6 | 13.1 | 2.8 |
| Upland Hardwoods | 881 | 2.50 | 1523.8 | 476.9 | 16.4 | 2.7 |
| Teak | 3364 | 9.56 | 1314.6 | 353.0 | 26.1 | 2.9 |
| Sal | 4251 | 12.08 | 1435.2 | 348.3 | 24.6 | 2.7 |
| Bamboo Forest | 567 | 1.61 | 2268.3 | 564.9 | 23.8 | 2.7 |
| Mangrove | 201 | 0.57 | 1734.3 | 280.8 | 26.6 | 2.5 |
| Miscellaneous forest | 22339 | 63.48 | 1679.8 | 374.5 | 23.0 | 2.7 |
| Western Ghat evergreen forest | 163 | 0.46 | 3111.3 | 368.7 | 25.4 | 2.4 |

*Source:* Forest types and area[12]

Changes in climate in the forest areas are assessed using the B2 scenario projections and FSI categories of forests. The temperature as well as rainfall means are obtained by considering all the grids of each forest type, occurring in different parts of India. In general, under the B2 scenario projections, the mean rainfall (and mean temperature) in areas under forest cover is somewhat higher than that in the non-forested areas. The increase expected in rainfall under the changed climate is also relatively larger for the forested areas, about 376 mm compared to the overall average of about 235 mm. The mean change in temperature, however, is not different from that in the non-forested regions. As expected, the changes in climate are not uniform across the different forest types - ranging from a large increase of more than 550 mm/year for hardwood and bamboo forests to a modest 220 mm for the colder Fir/Blue-Pine forests (Table 1).

The changes in temperature also show a striking pattern, with colder forests being subjected to a larger increase of about 3°C, compared to the Western Ghat evergreen forests, which on an average become warmer by only about 2.4°C, compared to the national average of 2.9°C under the B2 scenario. The changes under the more extreme A2 scenario are qualitatively similar to those described above, except that the magnitude of change is larger. Most of the forests show an increase of about 4°C with the northern temperate forests being subjected to about 4.6°C increase, while the Western Ghat evergreen forests show the least change of about 3.3°C.

*Impact of climate change on forest types*

A comparison of the extent of area that is likely to occur in each of the forest types under the present climate regime, and that under the two future climate scenarios reveals the magnitude of changes that are expected to take place in each of the forest types. The BIOME4 model was run for a total of 10,864 grid points (10 minute x 10 minute) located in the Indian region, using the CRU 10-minute climatology. Due do gaps in data related to soil parameter values, the model could assign vegetation types to only 10,429 of these grid points. As mentioned earlier, a comparison with the FSI database (available at a much finer resolution of 2.5 minute x 2.5 minute) allowed us to use the information from 35,190 FSI grids. There was a reasonable match between the forest types predicted by BIOME4 with the forest types assigned by FSI. Thus, tropical evergreen forests were seen in the southern Western Ghats and in the northeastern region, while the temperate forests were seen to occur in regions corresponding to Fir/Spruce/Deodar forests.

*Impacts on BIOME4 vegetation types in forested grids*

The number of forested grids under different vegetation types, each one accounting for at least 0.5% of the total grids, under the Current, A2 and B2 GHG scenarios is given in Table 2. Tropical Xerophytic Shrubland will undergo maximum change, where a reduction from 40% in the Current scenario to about 2% to 2.5% under A2 and B2 scenarios can be observed (Table 2), possibly due to an increase in rainfall. Tropical Evergreen Forest is likely to increase from about 3% of the grids under Current scenario to 35% under A2 and 21.5% under B2 scenario due to an increase in the rainfall and a moderate increase in temperature. Similarly, Tropical Savanna is projected to increase from about 4% under Current scenario to 26% under A2 and 18% under B2 scenario. However, the Warm Mixed Forests and the Tropical Semi-deciduous forests are likely to be impacted to only a moderate extent under the two scenarios.

**Table 2.** Number and percent of forested grids undergoing change in vegetation types under A2 and B2 GHG scenarios, compared to the Current (non-GHG) scenario

| Vegetation types | Current | | A2 | | B2 | |
|---|---|---|---|---|---|---|
| | No. of grids | % grids | No. of grids | % grids | No. of grids | % grids |
| Tropical xerophytic Shrubland | 14160 | 40.24 | 706 | 2.01 | 902 | 2.56 |
| Tropical deciduous forest/woodland | 9389 | 26.68 | 8141 | 23.13 | 14906 | 42.36 |
| Warm mixed forest | 4753 | 13.51 | 3210 | 9.12 | 3782 | 10.75 |
| Tropical semi-deciduous forest | 2790 | 7.93 | 1069 | 3.04 | 744 | 2.11 |
| Tropical savanna | 1549 | 4.40 | 9225 | 26.21 | 6485 | 18.43 |
| Tropical evergreen forest | 962 | 2.73 | 12309 | 34.98 | 7563 | 21.49 |
| Temperate conifer forest | 274 | 0.78 | 297 | 0.84 | 413 | 1.17 |
| Temperate sclerophyll woodland | 258 | 0.73 | 30 | 0.09 | 59 | 0.17 |
| Cool conifer forest | 234 | 0.66 | 22 | 0.06 | 74 | 0.21 |
| Evegreen taiga/montane forest | 221 | 0.63 | 55 | 0.16 | 73 | 0.21 |
| Cold mixed forest | 183 | 0.52 | 31 | 0.09 | 47 | 0.13 |

*Distribution of grids for each forest type under control and GHG scenario*

A more detailed assessment of the impact of climate change likely to be experienced by different forest types in India under the B2 scenario is shown as a matrix in Table 3. For each of the vegetation type (shown as rows in Table 3), the percentage of grids that are likely to shift to a different vegetation type are indicated in the corresponding columns; the diagonal elements show percentage of grids that are likely to remain unaltered. Thus, only about 6% of grids under the Tropical Xerophytic Shrubland will remain unchanged under the B2 scenario, with 59% grids changing into Tropical Deciduous Forests/Woodlands and about 32% changing into Tropical Savanna. Interestingly, though 8% of the grids under Warm Mixed Forests change into Temperate Conifer Forests, reciprocally, about 37% of the grids under Warm Mixed Forests are likely to change to Warm Mixed forests, highlighting the complex interplay of different factors governing the occurrence of forest types at different locations. Rather reassuringly, the existing Tropical Evergreen Forests are likely to remain so under the B2 scenario, and the increase in area under this forest type is seen to be due to shifts experienced by the Tropical Deciduous and Tropical Semi-deciduous forest types, as seen from Table 3. The diagonal entries in the table also illustrate the higher vulnerability of the cold and temperate forests compared to the tropical ones.

**Table 3.** Illustration of changes in forest types; number of grids under control scenario and % of grids under GHG scenario (B2) for dominant forest types

| Forest Types | No. of grids in Control scenario | % of Grids under each forest type under the GHG scenario (B2) | | | | | | | | | |
|---|---|---|---|---|---|---|---|---|---|---|---|
| | | TPXS | TPD/WL | WM | TPSD | TPS | TPEG | TMC | TMSW | CC | ET/M | CLDMX |
| TPXS | 14160 | 6 | 59 | 0 | 0 | 32 | 0 | 0 | 0 | 0 | 0 | 0 |
| TPDWL | 9389 | 0 | 54 | 0 | 1 | 8 | 35 | 0 | 0 | 0 | 0 | 0 |
| WM | 4753 | 0 | 16 | 58 | 7 | 0 | 9 | 8 | 0 | 0 | 0 | 0 |
| TPSD | 2790 | 0 | 1 | 0 | 7 | 0 | 91 | 0 | 0 | 0 | 0 | 0 |
| TPS | 1549 | 0 | 29 | 0 | 0 | 66 | 4 | 0 | 0 | 0 | 0 | 0 |
| TPEG | 962 | 0 | 0 | 0 | 0 | 0 | 100 | 0 | 0 | 0 | 0 | 0 |
| TMC | 274 | 0 | 8 | 37 | 6 | 0 | 41 | 5 | 0 | 0 | 0 | 0 |
| TMSW | 258 | 0 | 1 | 86 | 0 | 0 | 0 | 0 | 11 | 0 | 0 | 0 |
| CC | 234 | 0 | 0 | 94 | 0 | 0 | 0 | 0 | 2 | 2 | 0 | 0 |
| ET/M | 221 | 0 | 0 | 24 | 0 | 0 | 0 | 1 | 0 | 26 | 19 | 14 |
| CLDMX | 183 | 0 | 0 | 79 | 0 | 0 | 0 | 0 | 6 | 0 | 0 | 0 |

*TPXS: Tropical xerophytic shrubland*  
*TPD/WL: Tropical deciduous forest/woodland*  
*WM: Warm mixed forest*  
*TPSD : Tropical semi-deciduous forest*  
*TPS: Tropical savanna*  
*TPEG : Tropical evergreen forest*  
*TMC: Temperate conifer forest*  
*TMSW: Temperate sclerophyll woodland*  
*CC: Cool conifer forest*  
*ET/M: Evegreen taiga/montane forest*  
*CLDMX: Cold mixed Forests*

When analysis was carried out for the BIOME4 vegetation types, including the dominant 'miscellaneous' forest type for all grids (forested as well as non-forested), the results showed similar trends where Tropical Xerophytic Shrubland undergo large-scale reduction while Tropical Savanna and Evergreen Forests undergo expansion.

*Impact on FSI forest types in forested grids*

When the impact of projected climate on the forested grids based on the FSI forest types and area data is considered, the dominant Miscellaneous forest type (where no species dominates) distributed across different parts of India, occurring in different rainfall and temperature zones and dispersed in fragments of varying sizes, is projected to undergo large-scale changes with 75% of the grids in A2 and 67% of the grids likely to be subjected to change of forest type (Figure 2). The economically important forest types, such as *Tectona grandis*, *Shorea robusta*, Bamboo, Upland Hardwoods and Pine, are projected to undergo change. When Pine, Teak, Sal and Bamboo are considered, over 75% of the grids are projected to undergo change in A2 and B2 scenarios. The forest types, which are likely to undergo minimal or no change under both A2 and B2 scenarios are Western Ghats Evergreen, Semi-evergreen and Mangrove Forest types.

The overall impact of climate change could be assessed by considering the percentage of grids or area that show a change in the forest types under climate change scenarios. This change in the vegetation or forest type may be taken as an indicator of the vulnerability of the forest ecosystems to the projected climate change. Analysis for the 35,190 forested grids shows that 77% of the grids under A2 and 68% under B2 scenario are likely to undergo vegetation change. This indicates that well over half of the area under forests in India is vulnerable to the projected climate change under both A2 as well as the moderate B2 GHG scenarios. Similar trends were observed using IS92a scenario based HadRM2 climate outputs and BIOME3 model, which reported over 70% of all the grids projected to undergo change in vegetation types and potential vegetation [25,19].

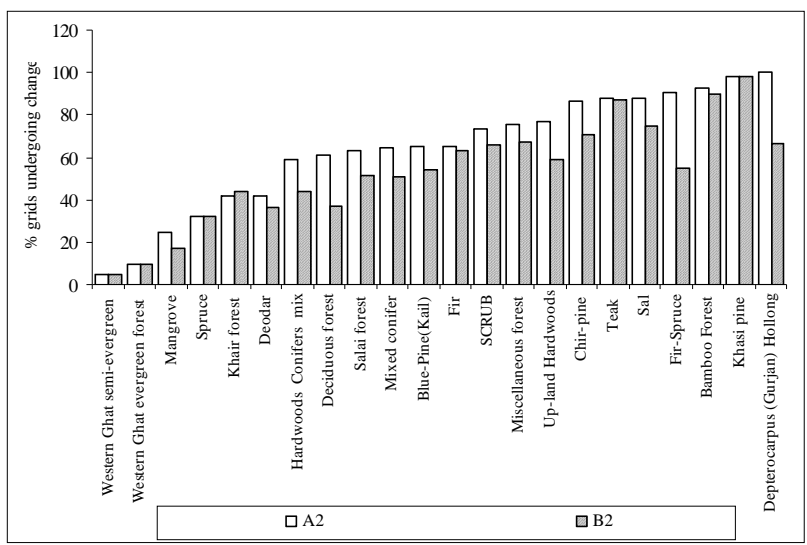

**Figure 2.** Percentage of grids under different forest types undergoing change in A2 and B2 GHG scenarios

A review of studies by IPCC[8] and Gitay et al.[11] has shown that forest biodiversity or the species assemblage is projected to undergo changes due to the projected climate change. Biodiversity is likely to be impacted under the projected climate scenarios due to the changes or shifts in forest or vegetation types (in 57 to 60% of forested grids), forest dieback during the transient phase, and different species responding differently to climate changes[8] even when there is no change in forest type. Climate change will be an additional pressure and will exacerbate the declines in biodiversity resulting from socio-economic pressures[11].

**Impact of climate change on net primary productivity**

The impact of climate change on net primary productivity (NPP in g $C/m^2$ per year) was estimated under the Current and GHG scenarios. The mean value of NPP is estimated to be 835 g $C/m^2$ per year under Current climate scenario for the forested grids. Under the A2 GHG scenario, a doubling of NPP is predicted, while the moderate B2 GHG scenario projects an increase of about 73% for the forested grids. NPP is projected to increase in all the forested grids mainly due to the $CO_2$ fertilization effect on forest ecosystems. The impact of climate change on NPP varies according to vegetation types (Figure 3).

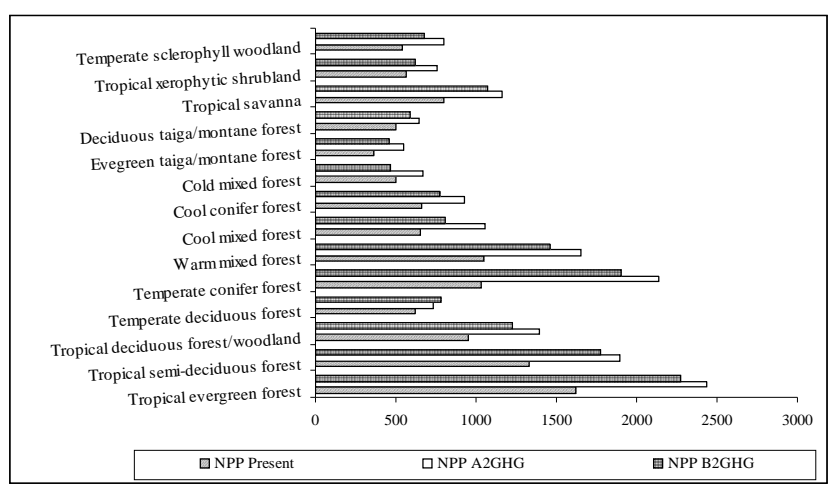

**Figure 3.** Climate impacts on NPP; % Forest biome-RCM grids subjected to change in NPP under GHG scenario over the current scenario under B2 Scenario

It can be observed from Figure 3 that among the dominant vegetation types (Tropical Xerophytic Shrubland, Tropical Deciduous Forest, Warm Mixed Forest and Tropical Semi-deciduous Forest), the NPP increases by 1.35 to 1.57 times under the GHG scenarios (A2 and B2) over the Current scenario NPP.

The NPP under Tropical Evergreen Forest increases by 1.5 times under the GHG scenarios. The rate of increase on NPP was lower for Cool Conifer Forest, Cold Mixed Forest and Temperate Deciduous Forest. Generally the rate of increase is higher for warmer vegetation types. A study, using the dynamic vegetation model, Hybrid-V4.1, which incorporated the patterns of nitrogen deposition and increasing $CO_2$ using the climate scenarios HadCM2 and HadCM3 for the period 1860-2100 predicted that vegetation carbon would increase by 290 GtC between 1860-2100 (compared to 600-630 GtC for the present day)[27]. Further, NPP is projected to increase from 45-50 GtC per year in the 1990s to about 65 GtC/year by 2080s in the HadCM2 scenario[27].

**Adaptation to climate impacts in forest sector**

The climate impact assessment made for Indian forest sector using regional climate model (HadRM3) outputs and BIOME4 vegetation model has shown that nearly 68 to 77% of the forested grids are likely to experience change, which includes loss of area under a given forest type as well as replacement by another type from the prevailing forest type by 2085. In other words over half of the vegetation is likely to find itself less optimally adapted to its existing location, making it vulnerable to adverse climatic conditions as well as to biotic stresses. Further the actual negative impact may be more than what is initially expected from the above description. This is because different species respond differently to the changes in climate[8]. Thus, one expects that a few species may show a steep decline in populations and perhaps even local extinctions. This, in turn, will affect the other taxa dependent on the different species (i.e., a 'domino' effect) because of the interdependent nature of the many plant-animal-microbe communities that are known to exist in forest ecosystems. This could eventually lead to major changes in the biodiversity. The positive impact of projected climate change, under the A2 and B2 scenario, is the projected increase in NPP. Thus, the projected climate impacts are likely to have significant implications for forest management in India.

Thus, climate change could cause irreversible damage to unique forest ecosystems and biodiversity, rendering several species extinct, locally and globally[8]. Forest ecosystems require the longest response time to adapt, say through migration and regrowth[10]. Further, a long gestation period is involved in developing and implementing adaptation strategies in the forest sector[28]. Thus there is a need to develop and implement adaptation strategies. Adaptation is adjustment in natural or human systems in response to actual or expected climatic stimuli and their impacts on natural and socio-economic systems. In the present study no attempt was made to assess or develop adaptation strategies due to the uncertainty involved in the climate impact assessment and the preliminary nature of the model outputs. However, several 'no regret' policies and forest management practices could be considered to address the impacts of climate change. Some examples of the 'no regrets' policies and practices are as follows:

- Incorporate climate concern in long-term forest policy making process
- Conserve forests and reduce forest fragmentation
- Expand Protected Areas and link them wherever possible to promote migration
- Promote mixed species forestry to reduce vulnerability
- Undertake anticipatory planting and assist natural migration through transplanting plant species
- Promote *in situ* and *ex situ* gene pool conservation
- Initiate forest fire management strategies.

Forest planning and development programmes may have to be altered to address the likely impacts of climate change and appropriately adopt various policy and management practices to minimize the adverse impacts and vulnerability. Adaptation strategies are also needed to ensure a proper balance between demand and supply of forest products.

**Future research needs; modelling, data and institutional**

The assessment of impact of climate change on forest ecosystems using regional climate model (HadRM3) for the period 2070 to 2100 and the equilibrium model BIOME4 has clearly demonstrated the possibility of; a large-scale shift in forest types in India with adverse implications for biodiversity and a large increase (nearly 70%) in net primary productivity of forest types with implications for biomass production and timber markets. Similar findings were obtained using IS92a based HadCM2 climate model outputs and BIOME3 model[26,19]. The trends of impacts could be viewed as robust but the magnitudes should be viewed with caution due to the uncertainties associated with projections of climate parameters using regional climate models as well as the limitations of the BIOME, an equilibrium model. Such a projected shift or change in forest types is likely to lead to large-scale forest dieback and loss of biodiversity. Forest ecosystems in India are already subjected to socio-economic pressures leading to forest degradation and loss, with adverse impacts on the livelihoods of the forest dependent communities. Climate change will be an additional pressure on forest ecosystems.

The Global Circulation Models are robust in projecting mean temperature at global level compared to their ability for making projections at regional level. The uncertainty involved in projections of precipitation changes is higher at global and particularly at regional level. The climate projections particularly the rainfall projections have high uncertainty and vary from model to model[29]. The BIOME is an equilibrium model and does not project the transient phase vegetation responses. The use of equilibrium and particularly the dynamic models is characterized by data limitations related to climate parameters, soil characteristics and plant physiological functions. Thus, the projections of impacts using the outputs of the current climate models as well as vegetation response models are characterized by high uncertainty. There is therefore a need to improve the reliability of climate projections at regional level as well as use of dynamic vegetation models. Data limitations need to be overcome by initiating studies to develop database on forest vegetation characteristics and plant functional types, plant physiological parameters, soil and water data and socio-economic dependence and pressures on forest ecosystems.

Development of adaptation strategies is constrained by uncertainty in the current projections of climate parameters as well as impact assessments. Further, there is a need for models where adaptation can be incorporated into impact models.

**Acknowledgements**

This study was carried out with support from the Joint Indo-UK Programme on "Impacts of Climate Change in India". We thank the Ministry of Environment and Forests and the British Government for the support. We would also like to thank Dr. Jed Kaplan for making the source code of BIOME4 available to us. We also thank IITM, Pune for providing climate projections and IIT, Delhi for providing soil and moisture data. We also benefited from the climate impact assessments carried out earlier under the National Communications project.